\documentclass[10pt]{myart}
\journal{Nuclear Physics B}
\usepackage{amssymb}
\usepackage{cite}
\usepackage{amsfonts}
\usepackage{epsfig}
\usepackage{color}
\newcommand{\be}{\begin{equation}}
\newcommand{\ee}{\end{equation}}
\newcommand{\bea}{\begin{eqnarray}}
\newcommand{\eea}{\end{eqnarray}}
\newcommand{\ba}{\begin{array}}
\newcommand{\ea}{\end{array}}
\newcommand{\nn}{\nonumber}
\newcommand{\set}[1]{\mbox{${\mathbb{#1}}$}}
\renewcommand{\Im}{\mbox{Im }}
\newcommand{\pref}[1]{(\ref{#1})}
\begin{document}
\begin{frontmatter}
\title{On Witten's global anomaly for higher \\ SU(2) representations\thanksref{thanksDOE}}
\thanks[thanksDOE]{This work is supported in part by funds provided by the U.S.
Department of Energy (D.O.E.) under cooperative research agreement
DF-FC02-94ER40818.}
\author{O. B\"ar}
\address{Center for Theoretical Physics, Laboratory for Nuclear Science\\
and Department of Physics,\\
Massachusetts Institute of Technology (MIT),\\
Cambridge, Massachusetts 02139, USA\\
email: obaer@lns.mit.edu\\
\vspace{0.5cm}
MIT preprint: MIT-CTP 3302
}

\begin{abstract}
The spectral flow of the overlap operator is computed numerically along a particular path in gauge field space. The path connects two gauge equivalent configurations  which differ by a gauge transformation in the non-trivial class of $\pi_4(SU(2))$. The computation is done with the SU(2) gauge field in the fundamental, the 3/2, and the 5/2 representation. The number of eigenvalue pairs that change places along this path is established for these three representations and an even-odd pattern predicted by Witten is verified. 

\end{abstract}
%\begin{keyword} %Anomalies in quantum field theory; Chiral gauge theories; Lattice gauge theory
%\PACS 11.15.Ha; 11.30.Rd
%\end{keyword}

\end{frontmatter}

\newpage
%
%========================================================================
\section{Introduction and summary of results}
%========================================================================
%
In 1982 Witten presented an argument that a chiral SU(2) gauge theory with a single left--handed doublet is mathematically inconsistent \cite{Witten:fp}. His observation was that the sign of the fermion determinant for this theory cannot be defined to satisfy both gauge invariance and smooth gauge field dependence. This obstruction is known as Witten's global SU(2) anomaly. Only for 
an even number of doublets is the theory well--defined since in that case the sign does not matter.

Even though a chiral SU(2) gauge theory in the fundamental representation is anomalous, many of the higher representations are not. 
In terms of the diagonal generator $T_3$, normalized as usual such that $\mbox{tr}\,T_3^{\,2}=1/2$ in the fundamental representation, only those representations are anomalous for which 
\be\label{Wittens_criterion}
2\,\mbox{tr}\,T_3^{\,2}\,=\,\mbox{odd integer}\,.
\ee
Alternatively, labeling the SU(2) representations by their highest weight $j$, the 
anomalous representations are given by
\be\label{anomalous_reps}
j\,=\, 2l+\frac{1}{2}\,,\qquad l\,=\,0,1,2\ldots\,.
\ee
Besides the fundamental representation with $j=1/2$, also the representations with $j=5/2, 9/2,13/2$ etc.~suffer from Witten's anomaly. The representations with integer values of $j$ as well as the remaining half-integer representations are anomaly free.

This pattern is somewhat puzzling. Of course, everyone is familiar with the difference between integer and half-integer representations of SU(2). So one might anticipate the absence of Witten's anomaly for the integer-valued representations. But what causes only half of the half-integer representations to be anomalous? The distinction between real and pseudo-real representations cannot offer an explanation here, since all half-integer representations are pseudo-real. 

Let us briefly recall Witten's argument to see where the criterion \pref{Wittens_criterion} comes from. First of all, the set of SU(2) gauge transformations is not continuously connected, but rather falls into two disjoint homotopy classes. This is mathematically expressed by 
\be\label{pi_4}
\pi_4(SU(2))\,=\, \mathbb{Z}_2\,,
\ee
and it means that there exist topologically non-trivial gauge transformations which cannot be smoothly deformed to the identity.

Taking a non-trivial gauge transformation $g$ we can define the linear interpolation
\be\label{cont_path}
A_{\mu}(x,t)\,=\,(1-t)\, A_{\mu}(x) + t\, A_{\mu}^g(x)
\ee
between an arbitrary gauge field $A_{\mu}$ and its gauge transform $A_{\mu}^g\,=\,g(A_{\mu} + \partial_{\mu})g^{-1}$. The $t$-dependent gauge field \pref{cont_path} defines a smooth path in configuration space.

Next, consider this path as the background gauge field for the massless Dirac operator
\be
D\,=\, \gamma_{\mu} (\partial_{\mu} + A_{\mu})\,.
\ee
$D$ is anti-hermitian and anti-commutes with $\gamma_5$, so the eigenvalues of $D$ are purely imaginary and come in complex conjugate pairs. Furthermore, the spectra at $t=0$ and $t=1$ are identical because the gauge fields for these $t$-values are gauge equivalent.

\begin{figure}[t]
\begin{center}
\vskip 0.5 cm
\setlength{\unitlength}{0.9mm}
\begin{picture}(80,40)

\put(10,0){\line(0,1){40}}
\put(70,0){\line(0,1){40}}
\put(9,20){\line(1,0){62}}

\put(10,10){\circle*{1.3}}
\put(10,6){\circle*{1.3}}
\put(10,2){\circle*{1.3}}

\put(10,30){\circle*{1.3}}
\put(10,34){\circle*{1.3}}
\put(10,38){\circle*{1.3}}

\put(70,10){\circle*{1.3}}
\put(70,6){\circle*{1.3}}
\put(70,2){\circle*{1.3}}

\put(70,30){\circle*{1.3}}
\put(70,34){\circle*{1.3}}
\put(70,38){\circle*{1.3}}

\linethickness{0.5pt}
\bezier{500}(10,30)(32,26)(40,20)
\bezier{500}(70,10)(48,14)(40,20)

\bezier{500}(10,10)(32,14)(40,20)
\bezier{500}(70,30)(48,26)(40,20)

\bezier{1000}(10,34)(40,24)(70,34)
\bezier{1000}(10,6)(40,16)(70,6)

\bezier{1000}(10,38)(40,29)(70,38)
\bezier{1000}(10,2)(40,11)(70,2)

\put(6.5,-3){$t=0$}
\put(66.5,-3){$t=1$}
\put(39,16){$t_c$}
\put(4,39){$\lambda_i$}

\end{picture}
\vspace {0.3cm}
\caption{\label{fig0} \small The simplest example of how the eigenvalue flow as a function of the path parameter $t$ could look like. One eigenvalue pair crosses zero at $t_c$ and changes places at the end of the path.} 
\end{center}
\vspace{0.3cm}
\end{figure}
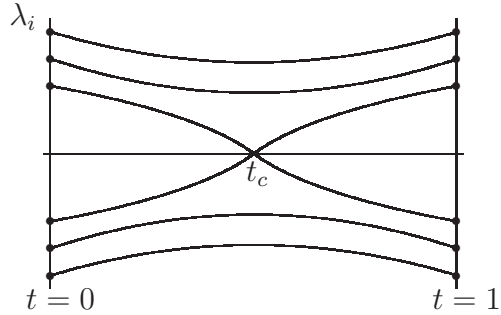

Witten proved, by invoking an Atiyah-Singer index theorem \cite{Atiyah:1971} for a certain five-dimensional Dirac operator (the fifth coordinate is essentially the path parameter $t$), that along the path \pref{cont_path} an odd number of eigenvalue pairs $\{\lambda(t), \lambda^*(t)\}$ cross zero and change places: 
\be\label{sign_flip}
\lambda_{i_k}\Bigg|_{t=0}\,=\,\lambda^*_{i_k}\Bigg|_{t=1}\,,\qquad k\,=\,1,\ldots,n\,,\qquad n\mbox{ = odd}\,.
\ee
The simplest scenario with only one eigenvalue pair changing places is sketched in figure \ref{fig0}. Note that the spectrum of $D$ is discrete because Witten considered space--time to be a four dimensional sphere and therefore to be compact.
%=========================
%: figure 1
%=========================
%=========================

Having established this result for the spectral flow, Witten showed 
that a chiral SU(2) gauge theory cannot be properly defined. The integration over the fermionic fields in the functional integrals for a theory with a single doublet of Weyl fermions yields, up to a sign, the square root of the determinant of the Dirac operator $D$. The sign of the square root is ambiguous and needs to be defined by hand. As a result of \pref{sign_flip}, any smooth and local sign definition leads to 
\be\label{sqrt}
\sqrt{\det D(A_{\mu}(t=0))} \,=\, (-1)^n\,\sqrt{\det D(A_{\mu}(t=1))}\,.
\ee
Since $n$ is odd gauge invariance is spoiled and the theory is ill-defined. 
%This is known as Witten's global anomaly. 
%It is called a global anomaly since the crucial result \pref{sign_flip} for the eigenvalues only holds if the gauge transformation $g$ belongs to the non-trivial homotopy class of $\pi_4(SU(2))$. It neither holds for infinitesimal nor finite gauge transformations in the trivial homotopy class. 

So far the discussion was restricted to the fundamental representation of SU(2). 
It turns out that the integer $n$ in \pref{sign_flip} depends on the representation. The Atiyah-Singer index theorem predicts $n$ to be odd only for the representations with $2\,\mbox{tr}\,T_3^{\,2}$ being an odd integer. For all other representations $n$ is even and gauge invariance can be preserved.

%So besides a non-trivial $\pi_4$ of the gauge group, the integer $n$ in \pref{sign_flip} is crucial for the presence or absence of Witten's anomaly. 
Witten's anomaly therefore hinges on the the integer $n$.
It is fairly simple to show that the representations with integer-valued $j$ are anomaly free, even without the Atiyah-Singer index theorem. Since the integer-valued representations are real one can show that each eigenvalue of $D$ is twofold degenerate.  Hence the exponent of the sign in \pref{sqrt} is $2n$ (or $4n,6n$ etc.)  and the sign automatically vanishes, independently of $n$.

The same argument cannot be applied to the half-integer representations, which are pseudo-real. 
The anomaly pattern for the half-integer representations is therefore directly tied to $n$ itself. This  leads us to the following question: What is the value of $n$ as a function of $j$ that causes only half of the half-integer representations to be anomalous?

At this point it is important to note that Witten uses the criterion \pref{Wittens_criterion} only to label the anomalous representations. He does not claim that the value of $2\,\mbox{tr}\,T_3^{\,2}$ coincides with the total number $n$ of eigenvalue pairs that change places along the path. In view of the numbers for the first three half-integer representations,
\be
\ba{lcl}
j\,=\,\frac{1}{2}: & \hspace{0.3cm}&2\,\mbox{tr}\,T_3^2\,=\, 1\,,\nn\\[1.4ex]
j\,=\,\frac{3}{2}: & &2\,\mbox{tr}\,T_3^2\,=\, 10\,,\\[1.4ex]
j\,=\,\frac{5}{2}: & &2\,\mbox{tr}\,T_3^2\,=\, 35\,,
\ea
\ee 
it is indeed hard to believe that 10 and 35 eigenvalues change places for the 3/2 and 5/2 representation, respectively. In fact, the mod-two index theorem for skew--adjoint elliptic operators \cite{Atiyah:1971} used by Witten does not predict any details about the spectral flow of the Dirac operator. All it predicts is whether $n$ is even or odd, not $n$ itself. 

It is worth mentioning that there is another way to study Witten's anomaly by embedding the gauge group SU(2) in a larger group, SU(3) for example \cite{Elitzur:1984kr,Klinkhamer:1990eb}. By doing so one also finds the representations with $j$ given in \pref{Wittens_criterion} to be anomalous \cite{Zhang:1987pw,Bar:2000qa}. However, also this method lacks any insight concerning the spectral flow of the Dirac operator. The value of $n$ remains unknown. 

In this paper we compute $n$ for the 3/2 and 5/2 representations. Our method is conceptually fairly simple. We formulate the path \pref{cont_path} on a discrete space-time lattice and compute numerically the spectral flow along this path for a suitable lattice Dirac operator. Suitable here means that the Dirac operator should satisfy the Ginsparg-Wilson relation \cite{Ginsparg:1981bj} in order to preserve exact chiral symmetry even for finite lattice spacing. The feasibility of this method has already been demonstrated for the fundamental representation \cite{Bar:1999ka} (see also \cite{Neuberger:1998rn}). There the spectral flow was found to look roughly as in figure \ref{fig0}, where $n$ is 1. 

We find $n$ to be 2 and 3 for the 3/2 and 5/2 representation, respectively. Hence the results for the lowest three half-integer representations  can be summarized by the simple relation
\be\label{my_conjecture}
n\,=\, j + \frac{1}{2}\,.
\ee
It is tempting to propose that \pref{my_conjecture} holds for all $j$. After all, \pref{my_conjecture} is the simplest relation one can imagine that explains the odd-looking anomaly pattern in terms of the spectral flow. Needless to say, a rigorous proof of \pref{my_conjecture} for all $j$ is beyond the scope of a numerical calculation and needs to be found elsewhere.

In addition to our results for $n$ we find some unexpected behavior for the spectral flow.
In order for an eigenvalue pair to change places, it must cross zero at least once along the path. The actual spectral flow for the higher representations, however, is more complicated. For the 3/2 representation we find 4 zero crossings, one eigenvalue pair crosses zero 3 times, a second one once. For the 5/2 representation it is even more complicated: In total 9 zero crossings are found. The results suggest that one eigenvalue pair crosses 5 times, another one 3 times and a third one once, even though our numerical approach cannot establish this beyond any doubt.

These results are obtained for one particular path. This is sufficient for determining $n$, since it is  path independent and only depends on whether the gauge transformation belongs to the non-trivial class of $\pi_4(SU(2))$ or not. In other words, it is a homotopy invariant. In contrast, there is a priori no reason why the number of zero crossings should be path independent. It is fairly possible that it depends on the specific starting and end point of the path or the details of the path itself.

To check for this the computation of the spectral flow has been repeated for different paths. Randomly generated gauge transformations $g$ in the non-trivial homotopy class have been used in the definition \pref{cont_path} of the path. For all paths studied we have always found  
1 zero crossing for the fundamental representation and 4 zero crossings for the 3/2 representation. In particular, we have never found the naively expected 2 zero crossings for the 3/2 representation. Unfortunately, the same check could not be carried out for the 5/2 representation for lack of computer resources.

Our results for the total number $k$ of zero crossings along the linear interpolation \pref{cont_path} can be summarized by 
\be\label{conjecture_2}
k\,=\,\left(j+\frac{1}{2}\right)^2\,=\, n^2\,.
\ee
Whether this relation is valid for all representations $j$ and whether it holds for all paths in gauge field space is unknown. Again, answering these questions is beyond the scope of a numerical study and this paper. 

Having already summarized our main results, the rest of the paper is organized as follows. 
In section 2 we give a homotopically non-trivial gauge transformation and construct the lattice analogue of Witten's path \pref{cont_path}. We also discuss briefly the relevant spectral properties of the overlap operator \cite{Neuberger:1997fp,Neuberger:1998wv}
which was used in the numerical computation. After describing some numerical details 
the results for the spectral flow are presented in section 3 and 4. Concluding remarks are made in section 5 and additional details can be found in the appendix.
%
%========================================================================
\section{Preliminaries}
%========================================================================
%
%
%==================================================
\subsection{A topologically non-trivial gauge transformation}
%==================================================
%
An explicit example of the non-trivial element of
$\pi_4$(SU(2)) has been constructed in \cite{Lueschernd1}. 
The main point in the construction is the observation that the group
SU(3) is a non-trivial SU(2) bundle over $S^5$. Starting from local
trivializations of that bundle on the upper and lower half-sphere of $S^5$, the
transition function on the equator is a homotopically non-trivial 
mapping from $S^4$ into SU(2). In terms of the Pauli matrices $\sigma_j, j
= 1,2,3,$ the final result in \cite{Lueschernd1} can be written as
\bea\label{map_stereo_coord}
g(n) & = & \frac{\xi_0}{1+n_4^2}\sigma_0 +
i\sum_j \frac{\xi_j}{1+n_4^2}\sigma_j,
\eea
where $\sigma_0$ denotes the two dimensional identity matrix and 
$n$ is an element of $S^4 = \{n\in \set{R}^5, n^2 =1\}$. The
functions $\xi_{\mu}$ are given by
\be\label{def_xi}
\begin{array}{lcl}
\xi_0(n) & = & 2n_4,\\
\xi_1(n) & = & 2(n_0n_2 + n_1n_3),\\
\xi_2(n) & = & 2(n_0n_3 - n_1n_2),\\
\xi_3(n) & = & n_0^2 + n_1^2 -n_2^2 -n_3^2.
\end{array}
\ee
From this definition it follows immediately that
\bea\label{xi_squared}
\sum_{i=1}^3 \xi_i^2 = \left(\sum_{i=0}^3 n_i^2\right)^2. 
\eea
Hence $\xi=(\xi_1,\xi_2,\xi_3)$ is an element of $S^2$ if
$\tilde{n}=(n_0,n_1,n_2,n_3)$ is an 
element of $S^3$. Using \pref{xi_squared} one can easily check that $g(n)$ is indeed an element of SU(2). 

The coordinates $n$ in \pref{map_stereo_coord} are elements of $S^4$, used as stereographic coordinates of $\set{R}^4$. To introduce a Euclidean space time lattice we need to write $g(n)$ in terms of cartesian coordinates $x_{\mu}$. This is done by using the well-known relations 
\bea
n_{\mu} \,=\,\frac{2x_{\mu}}{x^2 + 1},\qquad n_4\,= \,\frac{x^2 - 1}{x^2 + 1}, \label{stereo_coord1}
\eea
between cartesian and stereographic coordinates. Here $x$ denotes the Euclidean distance of the space--time point  $x_{\mu}$. Note that in this convention the north pole $n_4=1$ corresponds to $x=\infty$. 

Using the formulae \pref{stereo_coord1} in \pref{map_stereo_coord} one finds the corresponding map from $\set{R}^4$ into SU(2) as
\bea\label{g(q)}
g(x) & = & \frac{x^4 -1}{x^4 +1} \sigma_0 + i \sum_j\frac{2\xi_j(x)}{x^4+1}
\sigma_j\,.
\eea
At the origin $g$ is
identical to $-\sigma_0$ and it becomes the identity in the limit $x\rightarrow\infty$.
Due to the fourth power in $x$ this limit is practically reached for moderate values $x$. 

The mapping $g$ has various  symmetry properties. For instance, it is
invariant under the transformation $x_{\mu} \rightarrow
-x_{\mu}$. In addition, there exists a transformation $T$ that
transforms $g$ into its inverse. This transformation is explicitly
given by
\be
T: \qquad
\ba{lcrclcr}
 x_0 & \rightarrow & - x_3, & \qquad& x_2 & \rightarrow & x_1, \\
 x_1 & \rightarrow & - x_2, & \qquad& x_3 & \rightarrow & x_0.
\ea
\ee
This unitary transformation satisfies $T^2=-1$ and we find 
\bea\label{symmetry_g}
g(Tx) & = & g^{-1}(x)
\eea
for the gauge transformation \pref{g(q)}.  
%
%=================================
\subsection{Definition of the path}
%=================================
%
Restricting the coordinates $x_{\mu}$ in \pref{g(q)} to the  points of a lattice defines a lattice approximation of the gauge transformation $g$. As long as the lattice spacing $a$ is small enough, this approximation will give rise to a spectral flow of a lattice Dirac operator equivalent to \pref{sign_flip}, even if the topological statement \pref{pi_4} does not hold for lattice gauge transformations \cite{Bar:1999ka}. When working on a finite lattice with periodic boundary conditions, one can simply 'cut' $g(x)$ for $x_{\mu}$ larger than the lattice size $L$ without spoiling  the spectral flow, provided $L$ is large enough. 

Let us construct the lattice analogue of Witten's path \pref{cont_path}. As the starting point we take the constant gauge field 
\bea\label{def_const_gauge_field}
V(x,\mu) & = & \exp\left(i\theta(\mu) \sigma_3\right)
\eea
with some angles $\theta(\mu)$ which will be specified later. 
%We write $V(x,\mu)$ even though the field  is independent of $x$.
Using $V(x,\mu)$ and its gauge transform $V^g(x,\mu) = g(x)V(x,\mu)g(x+a\hat{\mu})^{-1}$ we define the path 
\bea\label{path_numerical}
\ba{rcl}
U_t(x,\mu) & = & N(t)^{-1} \Bigm\{(1-t)V(x,\mu) + t
  V^g(x,\mu)\Bigm\},\\[1.4ex]
N(t) & = & \mbox{det}\Bigm\{(1-t) V(x,\mu) + t
  V^g(x,\mu)\Bigm\}.
\ea
\eea
The link $U_t(x,\mu)$ is well--defined for all values of $t$ as long as $V\neq V^g$. The normalization factor $N(t)$ ensures that $ U_t(x,\mu)$ is an element of SU(2). Notice that $U_t$ is a smooth function of $t$.

The path \pref{path_numerical} has a very useful symmetry property. 
Taking into account $N(t) = N(1-t)$ one finds for the link variables the relation
\bea
\ba{rcl}
U_t(x,\mu) & = & g(x)
\tilde{U}_{1-t}(x,\mu)g(x+a\hat{\mu})^{-1},\label{symmetry_links} \\[1.4ex]
\tilde{U}_{t}(x,\mu) & = & N(t)^{-1} \left\{(1-t)V(x,\mu) + t
  V^{g^{-1}}(x,\mu)\right\},
\ea
\eea
i.e. $U_t$ and $\tilde{U}_{1-t}$ are gauge equivalent. 
The gauge field $\tilde{U}_t$ differs from $U_t$ by using the inverse
gauge transformation $g^{-1}$ for the gauge transform of $V$. 

Taking into account \pref{symmetry_g} and \pref{symmetry_links} it is not difficult to prove that the
spectrum $\sigma$ of the well-known Wilson-Dirac operator $D$ shows the symmetry property
\bea\label{symmetry_property}
\sigma\left(D(U_t)\right) & = & \sigma\left(D(U_{1-t})\right)\,.
\eea
It is therefore sufficient to compute the spectral flow along the path for either $0\leq t\leq1/2$ or  $1/2\leq t \leq1$. This, of course, is beneficial in a numerical computation. 

So far we have restricted the definitions to the fundamental representation of SU(2). It goes almost without saying that \pref{g(q)} and \pref{path_numerical} immediately define corresponding gauge transformations and gauge field paths for the higher group representations by applying
\be
\ba{rcl}
g(x)\; & \longrightarrow &\; R_j\left[g(x)\right]\,,\\[0.2ex]
U_t(x,\mu)\; & \longrightarrow & \;R_j \left[ U_t(x,\mu)\right]\,,
\ea
\ee
where $R_j$ denotes the SU(2) matrix for the representation $j$. The symmetry property \pref{symmetry_property} for the spectrum is valid for all representations.  The explicit formulae for $R_{3/2}$ and $R_{5/2}$ can be found in  appendix \ref{appendix_A}.
%
%==================================================
\subsection{Spectral properties of the overlap operator}
%==================================================
%
The overlap operator \cite{Neuberger:1997fp,Neuberger:1998wv} proposed by Neuberger is an explicit example of a Dirac operator that satisfies the Ginsparg-Wilson relation \cite{Ginsparg:1981bj}. It therefore allows the definition of chiral fermions on the lattice which is necessary in the present context.\footnote{For an introduction to exact chiral symmetry and chiral fermions on the lattice the reader may consult the overviews \cite{Niedermayer:1998bi,Luscher:1999mt,Neuberger:1999ry}, for example.}

In terms of the usual Wilson--Dirac operator $D_w$ the overlap operator is given by 
\bea\label{def_Neub_op}
aD & = & 1-A(A^{\dagger}A)^{-1/2}\,,\\[0.2ex]
 A & = & 1 - aD_w\,.
\eea
The eigenvalues $\lambda_i$ of $aD$ lie on a unit circle around 1 in
the complex  plane. In addition, the eigenvalues come in complex conjugate pairs
$\lambda_i,\lambda_i^*$, because the overlap operator satisfies the hermiticity relation
\be\label{gamma5_hermiticity}
D^{\dagger}\,=\,\gamma_5 D \gamma_5\,.
\ee 
If $\omega_i$ denotes an eigenvector with eigenvalue $\lambda_i$, the vector $\tilde{\omega}_i \,=\,\gamma_5 \omega_i$ is an eigenvector with eigenvalue $\lambda_i^*$. Since $D$ is a normal  operator, its eigenvectors can be chosen to form an orthonormal basis for the fermion fields. 

We are interested in the spectral flow of the overlap operator along the path defined in \pref{path_numerical}. 
The overlap operator $D(U_t)=D_t$ admits to choose eigenbases that depend smoothly on the path parameter $t$. Given such a basis $v_i(t)$,  Witten's result \pref{sign_flip} for the spectral  flow means that for an odd number of eigenvectors the following holds:
\bea\label{crossing_eigenvectors}
D_0 v_{i_k}(0) \,=\,\lambda_{i_k} v_{i_k}(0) \quad \Longrightarrow\quad D_1 v_{i_k}(1) \,=\,\lambda_{i_k}^* v_{i_k}(1)\,,\quad k=1,\ldots, n_j .
\eea
Let us assume that the spectrum is non-degenerate at the beginning of the path. This can always be achieved by choosing the starting gauge field accordingly, provided that there is no exact symmetry leading to a degeneracy of the spectrum. In that case we can conclude
\bea
v_{i_k}(0)\,=\,\omega_{i_k}\,,\qquad v_{i_k}(1)\,= \kappa \,\tilde{\omega}_{i_k}\,,\\[0.2ex]
\tilde{v}_{i_k}(0)\,=\,\tilde{\omega}_{i_k}\,,\qquad \tilde{v}_{i_k}(1)\,= \kappa' \,\omega_{i_k}\,,
\eea
with $\kappa,\kappa'$ being at most phase factors. In that sense the eigenvectors have interchanged along the path. In short, this result might be expressed in terms of the imaginary parts of the eigenvalues as \be\label{sign_flip_2}
\Im\lambda_{i_k}\Bigg|_{t=0}\,=\,-\Im\lambda_{i_k}\Bigg|_{t=1}\,,\qquad k\,=\,1,\ldots,n_j\,.
\ee
The integer $n_j$ is odd for the anomalous representations with $j=1/2,\,5/2$, and even for $j=3/2$. The subscript $j$ has been added to emphasize the dependence on the representation $j$.

Notice that $n_j$ depends only on the representation $j$ and the topology of the continuum gauge transformation $g$, i.e.~on the class of $\pi_4(SU(2))$ the gauge transformation belongs to. The integer $n_j$ does not depend on the particular gauge transformation $g$. It is also invariant under smooth deformations of the path \pref{cont_path} because it cannot change continuously.

In order for an imaginary part to change sign, one would expect it must become zero at least once along the path. Even though eigenvector mixing sometimes complicates this picture\footnote{see  also appendix \ref{appendix_B}.}, let us define
\be
k_j\,=\,\mbox{number of zero crossings}
\ee
along the path in the representation $j$. From \pref{sign_flip_2} one would conclude that $k_j \geq n_j$. As we will see later on, $k_j$ is indeed larger than its minimal value $n_j$ for the 3/2 and 5/2 representation.  

One important issue deserves a comment. Due to the presence of the operator $(A^{\dagger}A)^{-1/2}$  smooth $t$-dependence of the overlap operator is not obvious. And indeed, this is only guaranteed if the gauge fields along the path are smooth. More precisely, a differentiable $t$-dependence of $D$  --- and consequently for the eigenvectors and eigenvalues of $D$ --- holds if $A^{\dagger}A$ has no zero mode  along the path \cite{Hernandez:1998et}. This, however, can be achieved by choosing the lattice spacing $a$ small enough in the discretization of the continuum gauge transformation $g$.

The symmetry property \pref{symmetry_property} also holds for the spectrum of the overlap operator. For the eigenvalues as a function of $t$ it implies 
\be\label{symmetry_lambda}
\Im\lambda_i (t)\,=\,\pm\Im\lambda_i(1-t)\,.
\ee
Here the positive sign holds for eigenvalue pairs that do not change sign, while the negative sign applies to the $n_j$ pairs that satisfy \pref{sign_flip_2}. As a consequence, the imaginary part of those $n_j$ pairs must cross zero at $t=0.5$. Turning the argument around, any eigenvalue pair that crosses zero at $t=0.5$ is one of those $n_j$ pairs in \pref{sign_flip_2}. Hence, in order to determine $n_j$ it is sufficient to compute the spectral flow for a small neighborhood around $t=0.5$. This will be essential for our determination of $n_j$.
%
%==================================================
\subsection{Computing numerically the spectral flow}
%==================================================
%
Some general remarks concerning the numerical calculations seem to be appropriate before discussing the results. For notational convenience we set the lattice spacing $a=1$ from now on. 

The computation of the spectral flow was done in two steps. First we computed the lowest eigenvalues of $D^{\dagger}D$ using the Conjugate Gradient algorithm \cite{Bunk94a,Kalkreuter:1995mm}. This gives the absolute value $|\lambda_i|^2$ of the eigenvalues of $D$. 

Since they lie on a unit circle around one in the complex plane, their absolute value and imaginary part are related by simple trigonometry, leading to
\bea\label{relation_imag_abs}
\mbox{Im } \lambda_i& = & \pm |\lambda_i| \sqrt{1-\frac{|\lambda_i|^2}{4}}\,.
\eea
Recall that both signs occur because the eigenvalues come in complex conjugate pairs.
Our goal here is to detect zero crossings of the eigenvalues, which means zero crossings of their imaginary parts. By using \pref{relation_imag_abs} only we cannot unambiguously decide whether an imaginary part crosses zero because the information about the sign is lost.

A way to follow the sign along the path using the eigenvectors of $D^{\dagger}D$ is described in \cite{Bar:2000qd}. Without going into the details, we state here that the imaginary part of $\lambda_i$ can be expressed as a scalar product of properly chosen eigenvectors of $D^{\dagger}D$:
\be\label{Im_value_numerical}
\mbox{Im } \lambda_i\,=\, \langle \omega_{+,i}| D | \omega_{-,i}\rangle\,.
\ee
The brackets stand for the usual scalar product for the fermion fields and $\omega_{+,i}$, $\omega_{-,i}$, denote the chiral eigenvectors of $D^{\dagger}D$ with eigenvalue $|\lambda_i|^2$ (Note that $D^{\dagger}D$ commutes with $\gamma_5$). In addition, the phases of the eigenvectors can be fixed by making use of the reality properties of SU(2). As a result, a zero crossing of $\mbox{Im } \lambda_i$ is accompanied by a sign change for the matrix element in \pref{Im_value_numerical}, which is easily  computed numerically. This method was employed 
in the vicinities where the eigenvalues of $D^{\dagger}D$ are close to zero in order to establish whether the imaginary parts change sign or not.

%The eigenvectors can also be used to keep track of the spectral flow by computing the scalar product
%\be
%\langle \omega_{+,i}(t+\epsilon) | \omega_{+,j}(t)\rangle\,=\, \delta_{i j} - {\mbox{O }} (\epsilon)\,.
%\ee
%This is indispensable if two or more eigenvalues are close together.

Any practical implementation of the overlap operator needs an approximation for the inverse square root of $A^{\dagger}A$ in \pref{def_Neub_op}. We used a truncated expansion in  Chebyshev  polynomials \cite{Hernandez:1998et}. Depending on the lowest eigenvalue of $A^{\dagger}A$, which was computed numerically first along the path, the highest degree in the polynomial approximation was chosen up to 180.

A truncated Chebyshev expansion results in truncation errors for the eigenvalues as well as their imaginary parts. A second source of uncertainty is the numerical error.
Both errors, however, can be controlled analytically and strict error bounds can be derived for the eigenvalues as well as for their imaginary parts \cite{Bar:2000qd}.

Several times we were rather vague about the size of the lattice spacing $a$ and the lattice extension $L$ in order to capture the non-trivial topology of the continuum gauge transformation $g$. Some experimentation revealed that an $8^4$ lattice is sufficient for that purpose. All the results presented in the next section were obtained on a lattice of this size.

Finally, the angles $\theta(\mu)$ in the constant gauge field \pref{def_const_gauge_field} were set to values between $0.3\pi$ and $0.6\pi$. Different values were chosen for each direction to remove as much degeneracy of the spectrum as possible. The reason for having used a constant gauge field as the starting point of our path was to produce a gap in the spectrum. If we had used the classical vacuum configuration the lowest eigenvalues would have been very close to zero along the entire path. In that case checking for zero crossings would have required very small truncation and numerical errors and would have been computationally too time consuming. 
%
%========================================================================
\section{Numerical results}
\begin{figure}[p]
\begin{center}
\includegraphics*[scale = 0.45, trim=0 35 0 200]{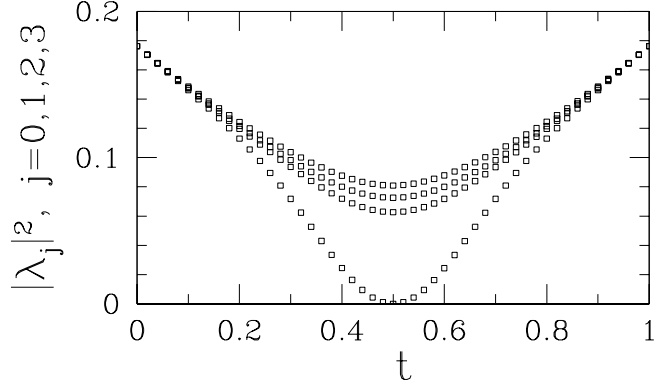}
\caption{\label{fig1} \small The lowest four eigenvalues of
 $D^{\dagger}D$ with SU(2) in the fundamental representation. The numerical and truncation errors are smaller than the size of the data points.} 
\vspace{0.3cm}
\end{center}
\end{figure}

%========================================================================
%
The spectral flow of $D^{\dagger}D$
for SU(2) in the fundamental representation is shown in figure \ref{fig1}. 
Since the computation for the fundamental representation is numerically not very demanding,
the eigenvalues were computed for the entire interval $[0,1]$. This serves as a check of the symmetry property \pref{symmetry_property}  of the spectrum, which is clearly visible.
%=========================
%: figure 1, 2 and 3
%=========================

\begin{figure}[p]
\begin{center}
\includegraphics*[scale = 0.45, trim=0 20 0 200]{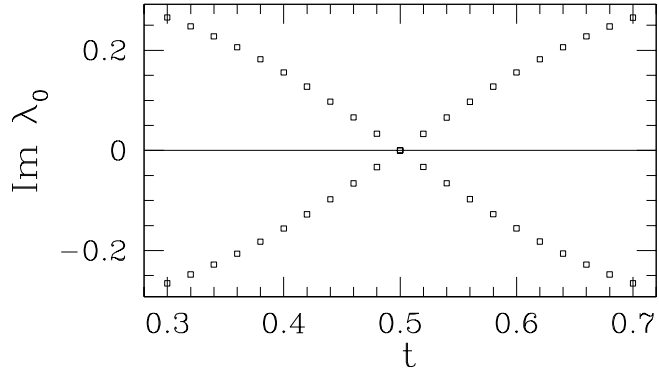}
\caption{\label{fig2}\small 
The imaginary parts of the lowest eigenvalue pair for SU(2) in the fundamental representation.
The eigenvalue pair crosses zero at $t=0.5$.
} 
\end{center}
\vspace{0.3cm}
\end{figure}

\begin{figure}[p]
\begin{center}
\includegraphics*[scale = 0.45, trim=0 20 0 200]{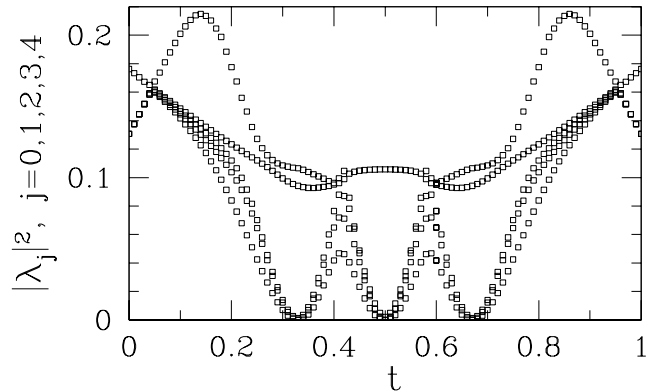}
\caption{\label{fig3}\small 
The lowest five eigenvalues of $D^{\dagger}D$ for the 3/2 representation. The data were taken for $0.5\leq t \leq1$ only and than reflected at $t=0.5$. 
%Around $t=0.33\,,0.5$ and $0.67$ three eigenvalues come close to zero and one needs to check for zero crossings.
} 
\vspace{0.3cm}
\end{center}
\end{figure}
\begin{figure}[p]
\begin{center}
\includegraphics*[scale = 0.45, trim=0 20 0 200]{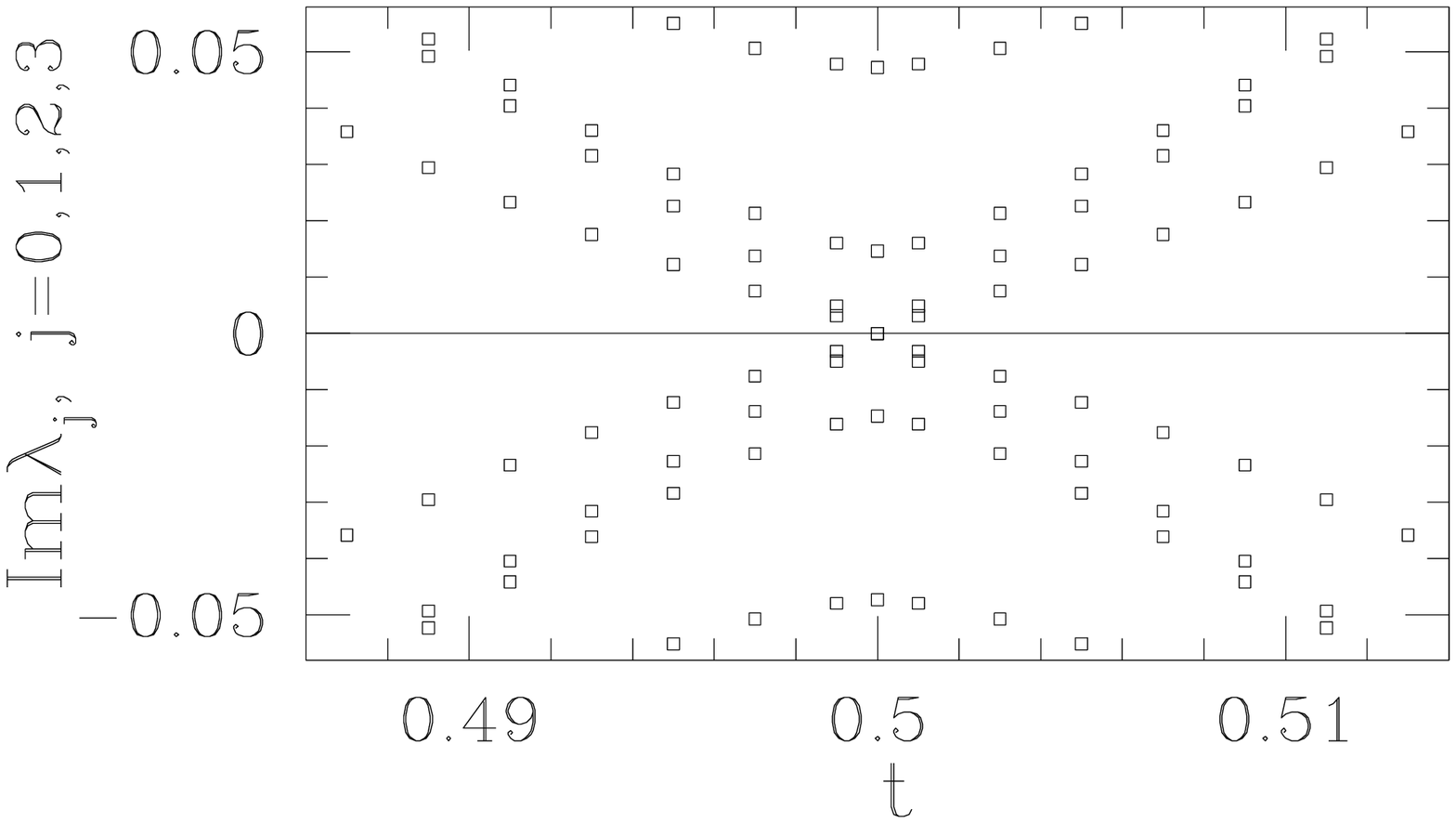}
\caption{\label{fig4}\small 
The imaginary parts of the lowest four eigenvalue pairs for $j=3/2$.
Two of the pairs cross zero at $t=0.5$. 
} 
\end{center}
\end{figure}
\begin{figure}[p]
\begin{center}
\includegraphics*[scale = 0.45, trim=0 20 0 200]{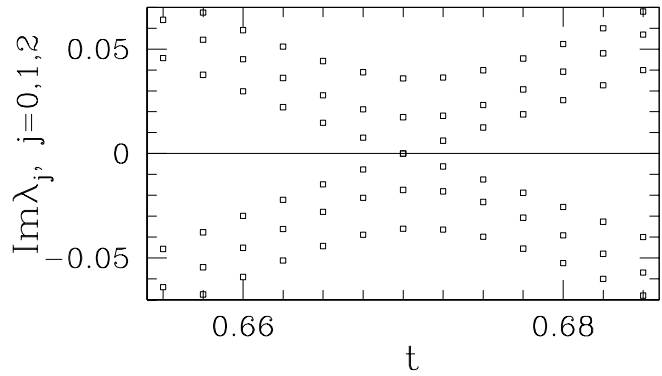}
\caption{\label{fig5}\small 
The imaginary parts of the lowest three eigenvalue pairs for $j=3/2$ around $t=0.67$.
One of the pairs crosses zero at $t= 0.67$. 
} 
\end{center}
\end{figure}
\begin{figure}[p]
\begin{center}
\includegraphics*[scale = 0.45, trim=0 20 0 200]{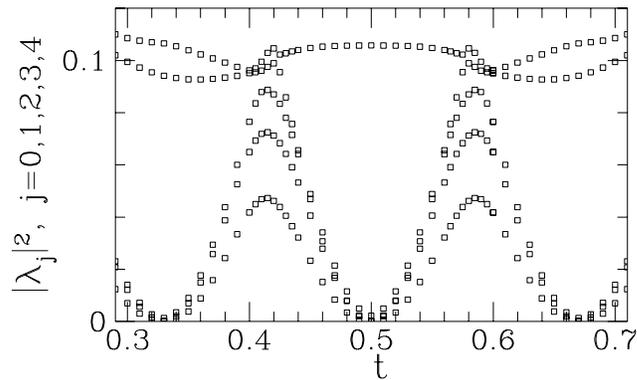}
\caption{\label{fig6}\small 
A magnification of figure~\ref{fig3}. The lowest eigenvalue is well separated from the higher ones between the two zero crossings at $t=0.5$ and $t=0.67$. This implies that the pair crossing zero at $t=0.67$ crosses zero again at $t=0.5$. 
%Because of the symmetry of the spectrum the lowest pair crosses zero three times along the path.
} 
\end{center}
\end{figure}

%===================================================================
As can be seen, the absolute value for all but one eigenvalue is never zero. 
Only the lowest eigenvalue becomes zero once at $t=0.5$.  Hence the imaginary part of all but the lowest eigenvalue cannot change sign along the path. 
The lowest eigenvalue pair, however, crosses zero at $t=0.5$ (figure \ref{fig2}) and we find
\be
\ba{l}
n_{1/2}\,=\,1\,,\\
k_{1/2}\,=\,1\,.
\ea
\ee
The results for $j=3/2$ are shown in figs.~\ref{fig3} to \ref{fig6} and can be summarized as follows: Two eigenvalue pairs cross zero at $t=0.5$ (figure \ref{fig4}), resulting in $n_{3/2}=2$.
This is in agreement with the Atiyah-Singer index theorem which requires $n_{3/2}$ to be even.
One of these pairs crosses zero again at $t = 0.67$ (figures \ref{fig5} and \ref{fig6}). Due to the symmetry property of the spectrum, this crossing also takes place at $t = 0.33$.  In total we find four zero crossings along the path. Hence
%=========================
%: figure 4,5  and 6
%=========================

%=========================
%=========================
%: figure 7,8 and 9
%=========================
\begin{figure}[p]
\begin{center}
\includegraphics*[scale = 0.45, trim=0 20 0 200]{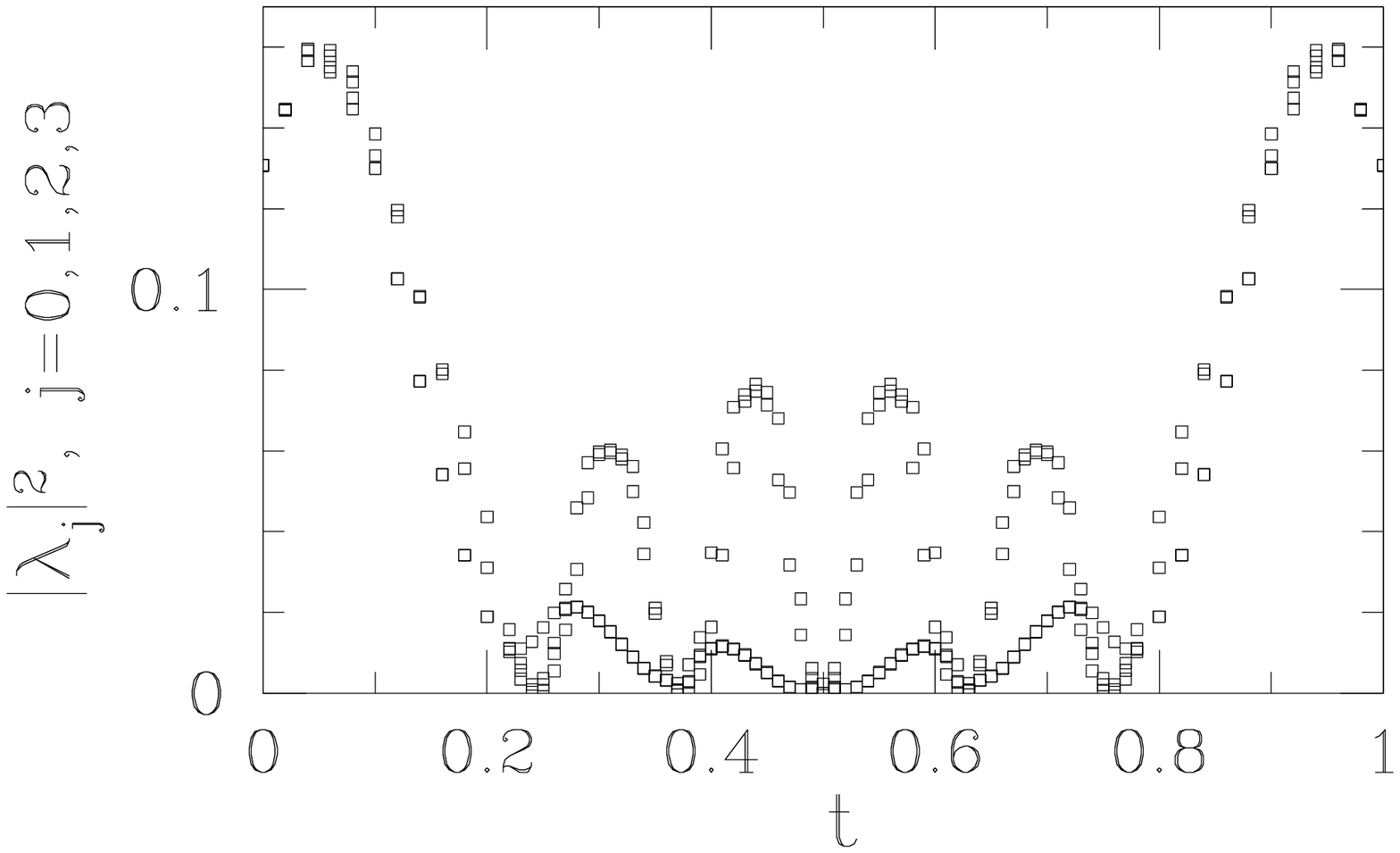}
\caption{\label{fig7}\small 
The lowest four eigenvalues of $D^{\dagger}D$ for the 5/2 representation. The data were taken for $0.5\leq t \leq1$ only and than reflected at $t=0.5$. 
} 
\end{center}

\end{figure}

\begin{figure}[p]
\begin{center}
\includegraphics*[scale = 0.45, trim=0 20 0 200]{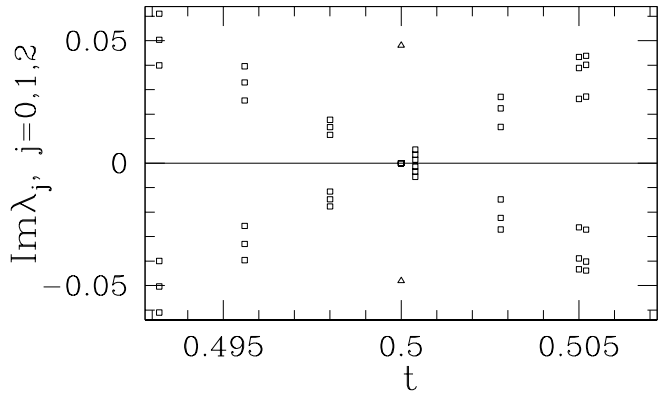}
\caption{\label{fig8}\small 
The imaginary parts of the lowest three eigenvalue pairs for $j=5/2$ around $t=0.5$.
All three pairs cross zero. The additional triangular data point at $t=0.5$ corresponds to the fourth eigenvalue. It is unequal to zero and therefore the fourth pair does not  cross zero.
} 
\end{center}

\end{figure}

\begin{figure}[p]
\begin{center}
\includegraphics*[scale = 0.45, trim=0 20 0 200]{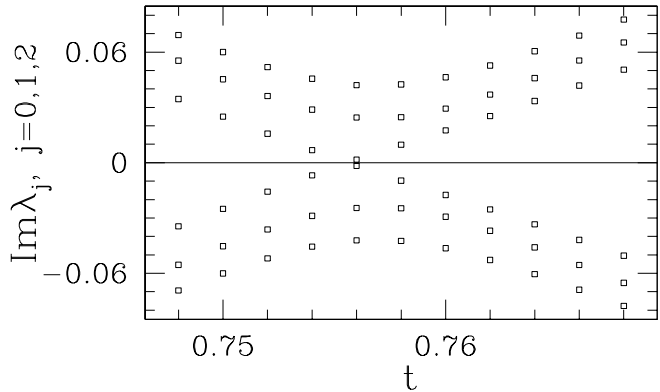}
\caption{\label{fig9}\small 
The imaginary parts of the lowest three eigenvalue pairs for $j=5/2$ around $t=0.76$.
There is one zero crossing at $t\approx 0.756$. 
}
\end{center}
\end{figure}
%=========================
%: figure 10
%=========================
\begin{figure}[t]
\begin{center}
\includegraphics*[scale = 0.45, trim=0 20 0 200]{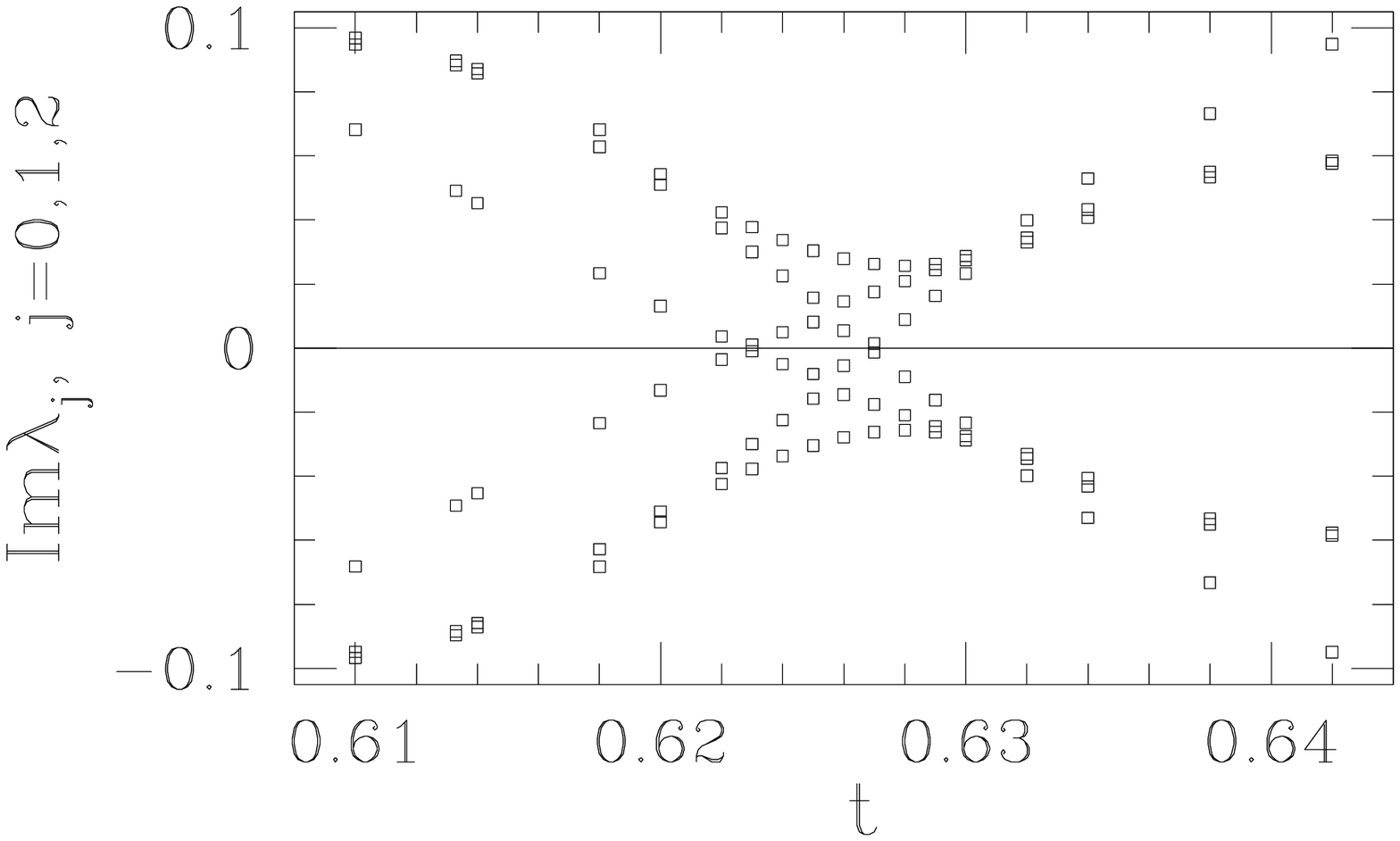}
\caption{\label{fig10}\small 
The imaginary parts of the lowest three eigenvalue pairs for $j=5/2$ around $t=0.63$. There is one zero crossing at $t\approx 0.623$ and a second one at $t\approx 0.627$. 
}
\end{center}
\vspace{0.3cm}
\end{figure}
%================================================================
\be
\ba{l}
n_{3/2}\,=\,2\,,\\
k_{3/2}\,=\,4\,.
\ea
\ee
Figures \ref{fig7} to \ref{fig10} show the results for the 5/2 representation. The spectral flow of the first four eigenvalues of $D^{\dagger}D$ (fig.~\ref{fig7}) has obvious similarities with the spectral flow for the 3/2 representation (fig.~\ref{fig3}). This time, however, there are five different $t$-values where eigenvalues are close to zero ($t= 0.24,\,0.37,\,0.5,\,0.63,\,0.76$).
At $t=0.5$ three pairs cross zero (fig.~\ref{fig8}), leading to $n_{5/2}=3$.  In addition, there is another zero crossing at $t\approx0.756$ and two zero crossings around $t=0.63$ (figs.~\ref{fig9} and \ref{fig10}). Due to the symmetry of the spectrum we finally find
\be
\ba{l}
n_{5/2}\,=\,3\,,\\
k_{5/2}\,=\,9\,.
\ea
\ee
As predicted by the Atiyah-Singer index theorem, $n_{5/2}$ is an odd integer.

Note that we are unable to rigorously establish whether one pair crosses zero five times, a second one three times and a third one only once. The eigenvalues are too close together to disentangle them properly (see fig.~\ref{fig10}).

This emphasizes the importance of the symmetry property \pref{symmetry_lambda} for the determination of $n_j$. Without making use of \pref{symmetry_lambda} we would need to keep track of each eigenvalue pair along the entire path, carefully counting every zero crossing. For the 5/2 representation this would have been absolutely impossible, at least for our particular path.

To summarize, the numerical results show the validity of e.q.~\pref{my_conjecture} for the first three half-integer representations. A proof for all $j$ is, of course,  beyond the scope of a numerical study like the one employed here.

The results for $k_j$ are slightly surprising. The number of zero crossings as a function of $j$ is given by equation \pref{conjecture_2}, which is larger than required by the results for $n_{j}$. 
Another peculiarity is that one eigenvalue crossing seems to pull down some of the higher eigenvalues. The second and third eigenvalue at $t\approx0.67$ in figure \ref{fig5}, for example, come very close to zero without actually crossing it. 

Before dwelling too much on these results one should make sure that they are not purely accidental, related only to the particular gauge transformation \pref{g(q)}. It seems possible that $k_j$ can be made smaller by choosing a different gauge transformation $g$ or by
smoothly deforming the path.

All the data shown so far have been obtained on an $8^4$ lattice. To check whether the peculiarities stem from the finite lattice size $L$ or the lattice spacing $a$,  additional computations for the 3/2 representation were performed on $10^4$ and $12^4$ lattices, varying the lattice size and the lattice spacing. The results are qualitatively the same as the ones shown here. The main difference is that the magnitude of the eigenvalues decreases for larger lattice sizes, as expected. 
%
%========================================================================
\section{Is $k_j$ a topological invariant?}
%========================================================================
% 
Strictly speaking, the answer to this question is no. The number of zero crossings surely is a path dependent quantity. By simply going back and forth along the path the number of zero crossings is trivially enlarged. By doing so, however, one can only increase $k_j$. More precisely, $k_j$ is multiplied by a factor which does not depend on the representation. Hence this trivial path dependence is easily "divided out" by normalizing to the fundamental representation and 
the question can be raised for the ratio $k_j/k_{1/2}$. We are, however, more interested in the question whether $k_j$ can be {\em decreased}, from four to two in the case of the 3/2 representation, for example.

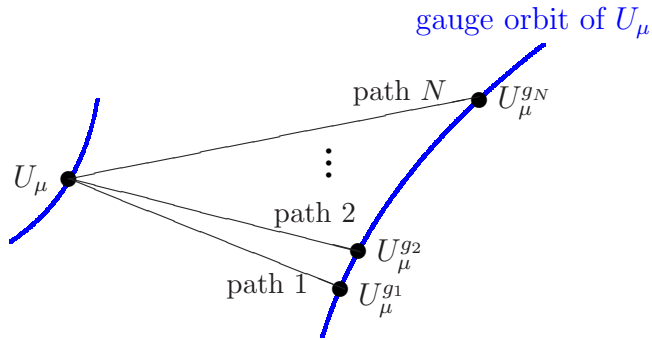
\begin{figure}[t]
\begin{center}
\setlength{\unitlength}{1.05mm}
\begin{picture}(60,40)

{\color{blue}
\linethickness{1pt}
\bezier{500}(32,0)(38,20)(60,37)
\bezier{200}(-7.5,12)(1.5,18)(3.5,30)
}
\put(44,39){\color{blue}gauge orbit of $U_{\mu}$}

\put(-7,19){$U_{\mu}$}
\put(0,20){\color{black}\circle*{2}}

\put(36.7,4){$U_{\mu}^{g_1}$}
\put(34.4,6){\color{black}\circle*{2}}
\put(0,20){\line(5,-2){35}}

\put(39,9.5){$U_{\mu}^{g_2}$}
\put(36.7,10.8){\color{black}\circle*{2}}
\put(0,20){\line(4,-1){37}}

\put(54.3,29){$U_{\mu}^{g_N}$}
\put(52,30){\color{black}\circle*{2}}
\put(0,20){\line(5,1){52}}

\put(33,23.5){\color{black}\circle*{0.6}}
\put(33,22){\color{black}\circle*{0.6}}
\put(33,20.5){\color{black}\circle*{0.6}}

\put(20,5.5){\small path 1}
\put(26,14.5){\small path 2}
\put(36,30){\small path $N$}
\end{picture}
\caption{\label{fig11} \small 
Using different homotopically non-trivial gauge transformations $g_i$ for the path \pref{path_numerical} amounts in changing its end point. 
}
\end{center}
\vspace{0.3cm}
\end{figure}

As a first step to answer this question it is certainly useful to check whether our result for $k_{3/2}$ persists for various non-trivial gauge transformations. Using different gauge transformations in the definition of the path means a variation of the end point, as sketched in figure~\ref{fig11}. If the four zero crossings we found above are a mere accident, related only to the particular gauge transformation $g$ in \pref{g(q)}, we should find examples for the spectral flow with only two zero crossings.

%%%%%%%%%%%%%%%%%%%%%%%%%%%%%
%:fig11
%
%%%%%%%%%%%%%%%%%%%%%%%%%%%%%
Topologically non-trivial gauge transformations can be generated numerically the following way. First of all one generates randomly a lattice gauge transformation $g$ with SU(2) elements $g(x)$ at each lattice point. This rough gauge transformation can be made smooth by slowly minimizing the functional
\be\label{Functional_F}
F[g]\,=\,\sum_{x,\mu}\,\mbox{tr}\bigg(\nabla_{\!\!\mu}\,g(x) \nabla_{\!\!\mu}\,g(x)\bigg)\,.
\ee
Here $\nabla_{\!\mu}$ denotes the usual forward difference operator. After sufficiently many smoothing steps the resulting gauge transformation represents a lattice discretization of a continuum gauge transformation which is either topologically trivial or non-trivial. The topological content can be checked by computing the spectral flow of $D$ for the fundamental representation. 
If $g$ is topologically non-trivial, there is exactly  one eigenvalue pair changing places, according to the findings in the previous section. 

Proceeding this way 10 suitable gauge transformations were generated and used to compute the spectral flow for the 3/2 representation. Suitable here means that the lowest eigenvalue of $A^{\dagger}A$ along the path was required to be not smaller than 0.01 for a given $g$. For eigenvalues that small the highest polynomial degree in the Chebyshev approximation for $(A^{\dagger}A)^{-1/2}$ has to be of the order of 300 to guarantee truncation errors as small as $10^{-4}$. This is already computationally quite demanding. Note that the randomly generated gauge transformations do not lead to \pref{symmetry_g} and the symmetry property \pref{symmetry_lambda} for the spectrum no longer holds. This doubles the amount of computer time necessary to compute the spectral flow.  

\begin{figure}[t]
\begin{center}
\includegraphics*[scale = 0.45, trim=0 20 0 200]{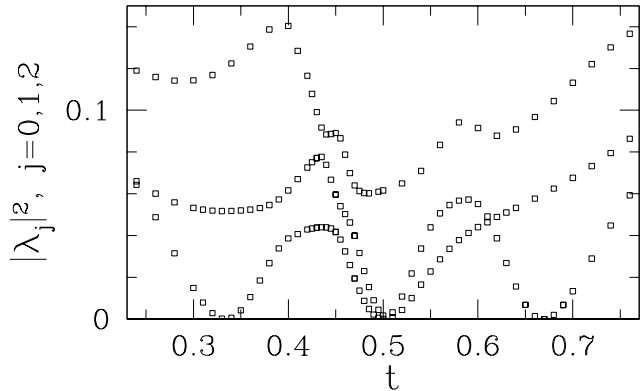}
\caption{\label{fig12}\small 
The lowest three eigenvalues of $D^{\dagger}D$ for the 3/2 representation, this time with a randomly generated gauge transformation $g$ in the definition \pref{path_numerical} of the path. 
Four zeros are clearly visible. 
} 
\end{center}
\vspace{0.3cm}
\end{figure}

Figure \ref{fig12} shows a typical example of the results. Plotted are again the eigenvalues of $D^{\dagger}D$, and the plot should be compared with figure \ref{fig6}. 
As anticipated, the spectrum is no longer symmetric under reflections at $t=0.5$. However, four zeros are visible. A direct computation of the imaginary parts shows indeed four zero crossings. A subtlety concerning fig.~\ref{fig12} is discussed in appendix \ref{appendix_B}.

Note that the second and third eigenvalue are well separated from the lowest one at $t\approx0.33$ and $0.67$. The "almost-zeros" in figure \ref{fig6} are therefore just a feature of the particular gauge transformation \pref{g(q)}. 

%=========================
%: figure 12
%=========================
%=========================

Qualitatively the same results have been found for all 10 randomly generated gauge transformations. Even though the spectral flows differ in the details for each gauge transformation, all cases show four zero crossings. No example with only two zero crossings has been found.

The spectral flow was also computed using different starting points for the path. Different constant gauge fields as well as a constant gauge field multiplied with a smooth but random gauge field were used. As before, no example with only two zero crossings has been found. 

It would have been desirable to perform the same checks for the 5/2 representation to see whether we always find nine zero crossings. Unfortunately, this test could not be carried out with the computer resources available to us. 

The number of zero crossings is remarkably stable under the path variations we have performed. However, one should note that we had to restrict ourselves to a rather special subset of gauge fields for the starting point of the path. In order to generate a gap for the spectrum, we have always chosen our starting field to be constant, except for small random fluctuations. It is possible that the number of zero crossings along the path changes for completely different starting fields.  

But even with its short comings, the numerical study of this section clearly demonstrates that the result $k_{3/2}=4$ is not connected to the analytically constructed gauge transformation \pref{g(q)} and its particular symmetry property. Whether  the number of zero crossings is of topological origin and given by a relation like the one suggested in \pref{conjecture_2} cannot be answered by a numerical study.  A possible strategy to show this analytically might be the embedding method introduced in \cite{Elitzur:1984kr}.
%
%========================================================================
\section{Concluding remarks}
%========================================================================
%
The main motivation for this work was to better understand Witten's criterion \pref{Wittens_criterion}  for the anomalous representations. Recall that the integer $2\,\mbox{tr}\,T_3^{\,2}$ takes the values $1, \,10$ and $35$ for the $1/2,\,3/2$ and $5/2$ representation, respectively. With the results presented  here we conclude that $2\,\mbox{tr}\,T_3^{\,2}$ is neither the number of zero crossings nor the number of eigenvalue pairs  that change places along the path. A connection between $n_j$ and $2\,\mbox{tr}\,T_3^{\,2}$ is obviously more subtle.

Witten's global SU(2) anomaly is widely known as an obstruction which forbids a chiral SU(2) gauge theory with an odd number of Weyl fermions in the fundamental representation. 
That higher representations exist which are anomaly free has not caught as much attention.
However, it remains to be proven that chiral gauge theories based on these representations have indeed no global anomalies when formulated on the lattice. This is not as obvious as in the continuum. Additional difficulties are imposed by a non-zero lattice spacing, a finite volume with some specific boundary conditions and the locality requirement  for the overlap operator \cite{Hernandez:1998et,Luscher:1999un}.

The integer-valued representations can be shown to be anomaly free also on the lattice \cite{Suzuki:2000ku}. Since these representations are real, the Dirac operator has an extra symmetry leading to a twofold degeneracy of the spectrum. This can be used to establish the absence of a global anomaly.

The pseudo-real half-integer representations are more difficult and the same argument cannot be applied. However, the lack of such an argument  does not a priori rule out an "accidental" degeneracy for some half-integer representations. Our numerical results explicitly show that there is no degeneracy for the 3/2 representation. A proof that a lattice chiral gauge theory based on this representation does not suffer from a global anomaly is still missing.
%
%========================================================================
\section*{Acknowledgments}
%========================================================================
%
I would like to thank M.~L\"uscher for many discussions on global anomalies. He also provided me 
with copies of his C-programs for the Conjugate Gradient algorithm and the random number generator ranlux. Many thanks go to I.~Campos for helping me with the numerics. 
Discussions with E.~Farhi,  O.~Philipsen and B.~Svetitsky are gratefully acknowledged.

%This work was supported in part by funds provided by the U.S. Department of Energy (D.O.E.) under cooperative research agreement DF-FC02-94ER40818.

{\section*{Appendix}
%
%========================================================================
\begin{appendix}
%========================================================================
%
\section{The 3/2 and 5/2 representation of SU(2)}\label{appendix_A}
The group SU(2) in the fundamental representation is the set of $2\times 2$ matrices $U$ which satisfy 
\be
U^{-1}\,=\,U^{\dagger}\,,\qquad \det U\, = \,1\,.
\ee
These two conditions are satisfied by matrices of the form
\be
U\,= \left(\ba{rr}
u_{11} & u_{12}\\
- u_{12}^* & u_{11}^*\ea
\right)\,,\qquad |u_{11}|^2 + |u_{12}|^2 \,=\,1\,.
\ee
The first row of the matrix determines the
matrix $U$ and the matrix elements $u_{11}$ and $u_{12}$ can be
used to parameterize higher representations of SU(2). The 3/2
representation, for instance, is four-dimensional and the matrix $R_{3/2}[U]$ reads \cite{Damgaard:2001fg}
\bea
\left(\ba{cccc}
u_{11}^3 & \sqrt{3} u_{11}^2 u_{12} & \sqrt{3} u_{11} u_{12}^2 &
u_{12}^3\\
-\sqrt{3} u_{11}^2 u_{12}^* & u_{11}\big(|u_{11}|^2 -
2|u_{12}|^2\big) &  -u_{12}\big(|u_{12}|^2 - 2|u_{11}|^2\big)& 
\sqrt{3} u_{11}^* u_{12}^2 \\
\phantom{-}\sqrt{3} u_{11} \big(u_{12}^*\big)^2 & u_{12}^*\big(|u_{12}|^2 -
2|u_{11}|^2\big) & \phantom{-}u_{11}^*\big(|u_{11}|^2 - 2|u_{12}|^2\big) & 
\sqrt{3} \big(u_{11}^*\big)^2 u_{12} \\
-\big(u_{12}^*\big)^3 & \sqrt{3}u_{11}^* \big(u_{12}^*\big)^2& 
-\sqrt{3}\big(u_{11}^*\big)^2 u_{12}^*& \big(u_{11}^*\big)^3
\ea\right).\nn
\eea
The six-dimensional 5/2 representation  $R_{5/2}[U]$ is given by
\bea
\left(\ba{ccc}
u_{11}^5 & \phantom{oo}& \sqrt{5} u_{11}^4 u_{12} \\
-\sqrt{5} u_{11}^4 u_{12}^* & & u_{11}^3\big(|u_{11}|^2 -
4|u_{12}|^2\big) \\
\phantom{-}\sqrt{10} u_{11}^3 (u_{12}^*)^2 & & \sqrt{2} u_{11}^2 u_{12}^*\big(3|u_{12}|^2-
2|u_{11}|^2\big) )\\ 
-\sqrt{10} u_{11}^2 (u_{12}^*)^3 & & \sqrt{2} u_{11}
(u_{12}^*)^2\big(3|u_{11}|^2- 
2|u_{12}|^2\big) \\
\phantom{-}\sqrt{5} u_{11} (u_{12}^*)^4 & &
(u_{12}^*)^3\big(|u_{12}|^2 - 4|u_{11}|^2\big) \\
-(u_{12}^*)^5 & & \sqrt{5} u_{11}^* (u_{12}^*)^4 
\ea \right.\hspace{0.3cm} \ldots \nn
\eea

\bea
\ba{cc}
\sqrt{10} u_{11}^3 u_{12}^2 &  \sqrt{10} u_{11}^2 u_{12}^3 \\
-\sqrt{2}u_{11}^2 u_{12} \big(3 |u_{12}|^2 - 2|u_{11}|^2\big) & \sqrt{2}u_{11} u_{12}^2 \big(3 |u_{11}|^2 - 2|u_{12}|^2\big) \\
\phantom{-}u_{11} \big(|u_{11}|^4 -6|u_{11}|^2|u_{12}|^2 + 3|u_{12}|^4\big) & u_{12} \big(|u_{12}|^4 -6|u_{11}|^2|u_{12}|^2 + 3|u_{11}|^4\big)\\
-u_{12}^* \big(|u_{12}|^4 -6|u_{11}|^2|u_{12}|^2 +3|u_{11}|^4\big) & u_{11}^* \big(|u_{11}|^4 -6|u_{11}|^2|u_{12}|^2 + 3|u_{12}|^4\big)  \\
\phantom{-}\sqrt{2}u_{11}^* (u_{12}^*)^2 \big(3 |u_{11}|^2 - 2|u_{12}|^2\big) & \sqrt{2}(u_{11}^*)^2 u_{12}^* \big(3 |u_{12}|^2 - 2|u_{11}|^2\big) \\
 -\sqrt{10} (u_{11}^*)^2 (u_{12}^*)^3 & \sqrt{10} (u_{11}^*)^3 (u_{12}^*)^2 
\ea\nn
\eea

\bea
\hspace{2.5cm}\ldots \hspace{0.3cm} \left.
\ba{ccc}
 \sqrt{5} u_{11} u_{12}^4 &  \phantom{oo} & u_{12}^5  \\
-u_{12}^3\big(|u_{12}|^2 -4|u_{11}|^2\big) & &
\sqrt{5} (u_{11}^*) u_{12}^4\\
\sqrt{2} u_{11}^* u_{12}^2\big(3|u_{11}|^2-2|u_{12}|^2\big) & &
\sqrt{10} (u_{11}^*)^2 u_{12}^3\\
-\sqrt{2} (u_{11}^*)^2 u_{12}\big(3|u_{12}|^2-2|u_{11}|^2\big) & &
\sqrt{10} (u_{11}^*)^3 u_{12}^2\\
(u_{11}^*)^3\big(|u_{11}|^2 -4|u_{12}|^2\big) &  & \sqrt{5} (u_{11}^*)^4 u_{12}\\
 -\sqrt{5} (u_{11}^*)^4 u_{12}^*
& &(u_{11}^*)^5
\ea\right)\,.\nn
\eea
Consider elements infinitesimally close to the identity, i.e.
\be
u_{11}\,=\, 1\, +\, i\,\frac{\epsilon_3}{2}\,, \qquad u_{12}\,=\,
\frac{\epsilon_2}{2}\, +\,i\,\frac{\epsilon_1}{2}\,. 
\ee
Keeping terms linear in $\epsilon_k$ only, one easily finds the
SU(2) generators for the 3/2 and 5/2 representation, respectively. By exponentiating them again, one recovers the representation matrices given above. In fact that was the way they were constructed.

\section{Zero crossings and eigenvector mixing}\label{appendix_B}

%=========================
%: figure 13
%=========================

\begin{figure}[t]
\begin{center}
\includegraphics*[scale = 0.45, trim=0 20 0 200]{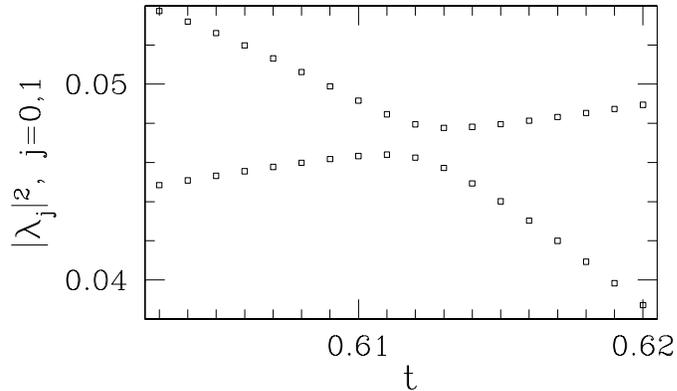}
\caption{\label{fig13}\small 
The lowest to eigenvalues for $t\approx 0.61$. the eigenvalues do not cross. 
}
\end{center}
\vspace{0.3cm}
\end{figure}
%=========================

Looking at figure \ref{fig11} one might be tempted to explain the sign changes of the lowest two eigenvalues by three zero crossings of one eigenvalue and one zero crossing of the other one. It is easy to smoothly connect the data points in figure \ref{fig11} with a pen according to this assumed eigenvalue flow. By doing so, however, the two lowest eigenvalues need to cross twice, once around $t=0.51$ and again around $t=0.61$. 

But this eigenvalue crossing does not take place. Figure \ref{fig13} shows the spectral flow around $t=0.61$. It is obvious that the eigenvalues do not cross as a function of $t$ (the numerical error in this plot is much smaller than the size of the symbols). Similarly, the two eigenvalues do not cross around $t=0.51$. Instead, one eigenvalue crosses zero four times while the second one is unequal to zero along the entire path.

This, however, does not contradict our previous result for $n_{3/2}$. Naively one might conclude from this that no imaginary part switches sign along the path. But one should remember that the sign change for the eigenvalues is defined in terms of the eigenvectors by eq. \pref{crossing_eigenvectors}, and not by the $t$-dependence of the eigenvalues. And \pref{crossing_eigenvectors} is still true with two imaginary parts changing sign, even though the lowest eigenvalue crosses zero four times. 

The reason we have not encountered this difficulty before is again the symmetry property \pref{symmetry_lambda}. This symmetry forces the eigenvalues to cross zero and each other at exactly the same value $t=0.5$, thus enabling us to extract $n_j$ by a local measurement of the spectral flow around $t=0.5$. Without this symmetry one has to track the eigenvectors along the entire path and use \pref{crossing_eigenvectors}, which is much more complicated.

The absence of a strict eigenvalue crossing is nothing new. It can also be found in QCD in the context of string breaking \cite{Philipsen:1998de,Knechtli:1998gf,Philipsen:1999wf,Knechtli:2000df}, for example. There one finds that the energy level $E_s$ for the string state, defined by a static $\bar{q}q$ pair connected by a flux tube, is smaller that the energy $E_{MM}$ of the two-meson state for $\bar{q}q$ separations $r$ smaller than the string breaking distance $r_b$. For $r>r_b$  the situation is reversed, the energy of the two-meson state is smaller than the energy of the string state. Naively one might expect a distance where $E_s$ equals $E_{MM}$. Mixing of the states for $r\approx r_b$, however, leads to a gap between the two eigenvalues of the Hamiltonian. But even though the energy levels do not cross, this does not contradict $E_s<E_{MM}$ for $r<r_b$ and $E_s>E_{MM}$ for $r>r_b$. This is analogous to what we discussed above for the spectral flow of the Dirac operator.

\end{appendix}
%: Bibliography
%========================================================================

%========================================================================
%

\end{document}